\documentstyle[epsfig,12pt]{article}
\begin{document}

\parskip 0.3cm
\begin{titlepage}
\begin{flushright}
CERN-TH/96-234\\
%hep-ph/9608
\end{flushright}

  \begin{centering}
 {\large {\bf Photoproduction of jets and the virtual structure of the photon}}\\
 \vspace{.4cm}
{\bf  D.de Florian \footnote{On leave of absence from Departamento de 
F\'\i sica, Universidad de Buenos Aires, Ciudad universitaria Pab.1 (1428) Bs.As., Argentina }}
\vspace{.05in}
\\ {\it Theoretical Physics Division, CERN, CH 1211 Geneva 23, Switzerland}\\
e-mail: deflo@na47sun05.cern.ch \\
 \vspace{.4cm}
{\bf  C.Garc\'\i a Canal} \\  
\vspace{.05in}
{\it Laboratorio de F\'{\i}sica Te\'{o}rica,
Universidad Nacional de La Plata,  \\ C.C. 67 - 1900 La Plata - 
Argentina}\\
\vspace{.4cm} 
{\bf R.Sassot} \\
\vspace{.05in}
{\it  Departamento de F\'{\i}sica, 
Universidad de Buenos Aires \\ 
Ciudad Universitaria, Pab.1 
(1428) Bs.As. - 
Argentina} \\
\vspace{.4cm}
{ \bf Abstract}
\\ 
\bigskip
{\small
We compute the ratio between the direct and the resolved photon components 
of single jet and dijet production in $ep$ collisions for the kinematical 
range covered by the most recent ZEUS data. We analyse the phenomenological 
consequences of different models for the  structure of virtual photons in 
these observables and compare them with the available  data. We also comment 
on the correlation between the so called $x_{\gamma}^{obs}$ and the `true' 
$x_{\gamma}$, that can be inferred from the data.}  
 \end{centering}
%\vspace{0.5in}
\vfill
\begin{flushleft}
CERN-TH/96-234\\
August 1996 \hfill
%\draft
\end{flushleft}
\end{titlepage}
\vfill\eject

\noindent{\large \bf Introduction:}\\

In recent years, different inclusive photon-photon experiments have contributed 
to unveil the parton content of photons in a program similar to that pursued 
for the proton. Several sets of parton distribution functions for the photon 
have been proposed and are periodically refined attaining increasing levels 
of precision. For a comprehensive review see references \cite{DG1,Sj1}.

The photoproduction of jets with large transverse energy at HERA \cite{Derr}, has opened 
the possibility of testing the gluon content of photons and the accuracy of these experiments
 allows also a clear discrimination 
between events generated by quasi-real and virtual photons \cite{utley}. These improvements 
make possible the testing of different hypothesis about the photon structure 
and its dependence on the virtuality scale, such as how the  hadronic 
component of the photon is supressed  at high virtuality, as it is usually 
expected. The details of this dependence were originally thought to be 
obtainable by means of a purely perturbative approach \cite{witt,uem}, at least for a 
restricted kinematical regime. More recently it has been analized  with non perturbative models 
for the hadronic structure of the photon at some definite energy scale \cite{glu,ss}. 
The above mentioned models  are the subyacent motivation of the most recent 
parton distribution parametrizations, whose  energy scale dependence is then 
driven by the inhomogeneus Altarelli-Parisi evolution equations \cite{witt}.

Both the perturbative and nonperturbative approaches generate photonic parton 
distributions which differ not only in the dependence on the virtuality scale 
but also in its quark and gluonic content. They can be compared and their consequences 
in different observables analysed.

Recently, the ZEUS collaboration have produced for the first time data on dijet 
photoprodution in $ep$ collisions for different values of the photon vituality
\cite{utley}. 
The preliminary data coming from this measurement  allows an interesting test 
for the current ideas on the virtual photon structure in the range of virtualities 
spanned between  $0.10$ and $0.55$ GeV$^2$. 

In this paper  we compute  the ratio between the resolved and the direct photon 
components of dijet photoproduction in $ep$ collisions,  already measured at 
HERA, using different models for the photon content.
In doing so, we take into account the non-trivial kinematical cuts inherent to 
the experimental data. Particularly, we show how  the magnitude and the 
dependence of this ratio in the photon virtuality is fairly reproduced by model 
dependent parametrizations for the photon, but not by the perturbative approach,
 at least at leading order, mainly due to the poor gluon content of this last approximation. 
We also determine the value of $x_{\gamma}^{threshold}$, the exact photon energy
 fraction threshold used to define the ratio, that provides the best agreement 
between theoretical estimates and experiment.
This procedure is necessary  due to the fact that the data points are obtained 
using  an experimentally defined threshold fraction, $x_{\gamma}^{obs}$, which 
is not straightforwardly related to the theoretically defined one. In doing this, 
we find a rather sensible agreement between the determinations coming from the
most realistic  parametrizations.

We also compute predictions for the same ratio but for single jet cross sections, 
as suggested in ref \cite{glu}, but in an integrated kinematical range, similar 
to that covered by ZEUS data analyses.
For this single jet cross sections, we find a clear increase in the ratio, i.e.
in the resolved component, due to low $x_{\gamma}$ contributions, which 
would suggest an increased sensitivity on the gluon component. However, the dependence  
on the virtuality scale is similar to that of dijet production.

In the following section we define the cross sections to be analysed, specifing 
the kinematical range over which these are integrated in order to be compared with 
the experimental data. Then, we make a short summary about the photonic parton 
distributions to be used and compare their main features. In the third section 
we show  theoretical estimates, compare them with the available data and discuss
 about the $x_{\gamma}^{threshold}$ choices that bring the best accord between them. 
Finally we sumarize our results and present our conclusions.\\

\noindent{\large \bf Jet Cross Sections:}\\

In leading order the differential cross section for two jet production in $ep$ collisions takes
a very simple form  when written in terms of the fraction of the photon energy intervening in the
hard process, $x_{\gamma}$, the fraction of the proton energy carried by the participating
parton, $x_p$, and that of the electron carried by the photon, $z$ \cite{DG0} 
\begin{equation}
 \frac{d\sigma}{dx_{\gamma}\,dx_{p}\, dz\, dp_{T}\,dP^2} = \tilde{f}_{\gamma /e}(z,P^2) f^{\gamma}(x_{\gamma},
Q^2,P^2)f^{p}(x_p,Q^2)\frac{d\hat{\sigma}}{dp_{T}} 
\end{equation}
Here, $p_T$ is the transverse momentum of the jets,  $P^2$ is the photon virtuality, and $Q^2$ 
is the relevant energy scale of  the proccess, in this case taken to be equal to $p_{T}^2$. 
$d\hat{\sigma}/dp_{T}$ represents the hard parton-parton and parton-photon cross sections
\cite{com}. 

The functions  $f^{\gamma}(x_{\gamma},Q^2,P^2)$ and $f^{p}(x_p,Q^2)$ denote the parton distribution
 functions for the photon and the proton, respectively. The first one reduces to $\delta(1-x_{\gamma})$ 
-the probability for finding a photon in a photon- for 
direct contributions, i.e. those in which the photon participates as such in the hard process.  
$\tilde{f}_{\gamma /e}(z,P^2)$ is the unintegrated Weizs\"acker-Williams distribution
\cite{man}
\begin{equation}
\tilde{f}_{\gamma /e}(z,P^2)=\frac{\alpha}{2\pi}\frac{1}{P^2}\frac{1+(1-z)^2}{z}
\end{equation}
which has been shown to be a very good approximation for the distribution of photons in the electron,
 provided the photon virtuality is much smaller than the relevant energy scale
\cite{glu}.

The differential cross section in eq.(1) can also be written in terms of the 
(pseudo) rapidities $\eta_1$ and $\eta_2$, which are constrained by the experimental 
settings, and the electron and proton energies $E_e$ and $E_p$ \cite{kram}.  Both pairs 
of variables are related to the energy fractions by  
\begin{eqnarray}
x_p &=& \frac{p_T}{2 E_p} \left( e^{\eta_1} + e^{\eta_2} \right) \nonumber \\
x_\gamma  &=& \frac{p_T}{2 z E_e} \left( e^{-\eta_1} + e^{-\eta_2} \right)
\end{eqnarray}
Kinematical restrictions constrain $x_\gamma$ to lay in the interval $[p_T^2 / x_p z E_e E_p , x_{\gamma}^{max}]$ , 
$x_p$ in $[p_T^2 / z E_e E_p x_{\gamma}^{max},1]$ and $z$ in  $[p_T^2/ E_e E_p,1]$.
In the ZEUS data there are also additional constraints involved such as those 
coming from  the cuts applied to the rapidities $-1.125\le \eta \le 1.875$, and  the ones 
for the Jacquet-Blondel variable $0.15\le y_{JB} \le 0.75$, which corresponds 
to  $0.20\le z \le 0.80$ \cite{Derr}.
In order to analyse the $P^2$ dependence of the  total cross section, 
the variables $x_{\gamma}$, $x_p$, $z$, and $p_{T}$, must be integrated taking into
account both sets of constraints.

Switching from the ($\eta_1,\eta_2$) plane (figure 1a) to the ($\tilde{x}_p,
\tilde{x}_\gamma$) (figure 1b)
where $\tilde{x}_p=x_{p}2E_{p}/p_{T}$ and $\tilde{x}_{\gamma}=x_{\gamma}2zE_{e}/p_{T}$, 
the triangle $ABC$ is mapped into the area $A'B'C'$ in figure 1b; the
triangle $ADC$ is also mapped into the same region, due to the indistinguishability of 
events related by an exchange of $\eta_1$ and $\eta_2$.
This implies that when integrating over $x_p$ and $x_{\gamma}$, two partonic
events must be considered for each point, the two partonic cross sections 
coming from the exchange of the Mandelstam variables $u$ and $t$ \cite{DG2},
\begin{equation}
\frac{d\hat{\sigma}}{dp_{T}}=\frac{d\hat{\sigma}}{dp_{T}}(s,u,t)+\frac{d\hat{\sigma}}{dp_{T}}(s,t,u)
\end{equation}    
For dijet process the $\eta$ cuts applied determine the regions $A'B'C'$ and $A'D'C'$
in the ($x_p,x_{\gamma}$) plane which have to be further constrained with the above mentioned kinematical 
cuts. For single jets events, due to the $\eta$ restrictions, only the areas where neither of the jets 
can be detected are excluded so the $x_{p},x_{\gamma}$ region is extended with the areas limited by
the prolongation of the curves $A'B'$ and $B'C'$
 
Experimentally, the ratio between the resolved and the direct contributions is 
defined  as the number of events with photon energy fractions $x_{\gamma}$  lower than 
certain threshold value $x_{\gamma}^{threshold}$ divided by the
 number of those events with greater energy fractions. Naively, one would expect the direct
contribution sharply peaked at  $x_\gamma=1$, i.e. the photon participating 
with all its energy in the hard process, so the most natural threshold for defining the ratio would be this,
 however, the detector resolution smears the distribution and also 
complicates the determination of $x_\gamma$, which is then approximated by a measurable
fraction $x_\gamma^{obs}$, defined in terms of the two highest  transverse energy jets.
Previous analyses have shown \cite{Derr} that a threshold of about $x_\gamma^{obs}=0.75$ 
provides a good discrimination between direct and resolved events, however it 
is not possible to determine to which value of the  $x_{\gamma}^{threshold}$ 
it corresponds \cite{utley}.

With these elements in mind, one is able to estimate the yield of an experimental determination 
of the ratio, been the set of photonic parton distributions and the 
value of $x_{\gamma}^{threshold}$ the main uncertainties left.\\

\noindent{\large \bf Photonic Parton Distributions:}\\

In recent years, several sets of parton distributions for the real photon 
have been proposed \cite{DG1,Sj1}. The scale dependence of these distributions is driven by the 
inhomogeneus Altarelli-Parisi evolution equations \cite{witt} with input distributions coming from 
either plausible dynamical assumptions or  phenomenological fits to  the photon structure function  
data, as for parton distributions in hadrons. 
For virtual photons a similar procedure can be followed, provided a dependence on the virtuality 
scale is somehow implemented.
In reference \cite{DG2}, for example, this dependence is introduced, for the quark distributions, 
by an interpolating factor multiplying  the real photon
parton distributions
\begin{equation}
r=1-\frac{\ln(1+P^2/P_c^2)}{\ln(1+Q^2/P_c^2)}
\end{equation}
where $P_c$ is a typical hadronic scale. For gluons, the factor is chosen to be the
square of the former. One can, for example, implement this approach in any of
the available real photon parton distributions.
Reference \cite{glu} proposes a decomposition between a perturbative and a nonperturbative component 
in the input parton distributions of virtual photons. The former
coming from the photon-photon box diagram, whereas the later is related to the parton content of pions. 
Both contributions are weighted by factors that guarantee a
smooth transition to the real photon description. 
A different decomposition is proposed in reference \cite{ss}, where the vector meson dominated 
contribution and the anomalous component  are multiplied
by certain dipole dampening factors, designed to take into account the $P^2$ dependence.

Altenatively, virtual photons offer another possibility which consists in 
obtaining the input parton distributions by a perturbative approach which also 
takes into account the dependence on the virtuality \cite{uem}. The caveat of this approach 
is that it is only applicable in a restricted kinematical region, ($\Lambda_{QCD}^2 << P^2 << Q^2$), 
where the photon exhibits hard pointlike behaviour and higher twist corrections
can be neglected. In leading order, the resulting parton distributions can then 
be approximated by 
\begin{equation}
q_{i}^{\gamma}(x,Q^2,P^2)\simeq\frac{\alpha}{2\pi}3 e_{i}^2 [x^2+(1-x)^2]\ln \frac{Q^2}{P^2} 
\,\,\,\,\,\,\,\,\,\,\,\,\,\,\,\,\,\,\,\,\,\,\,\,\,\,\,
g^{\gamma}(x,Q^2,P^2)\simeq 0
\end{equation}    
Next to leading order corrections to these distributions have also been computed
finding moderate corrections in the photon structure function but large ones 
for large 
values of $x$ \cite{uem}. 

In references \cite{glu,ss} it  has been noticed that most of the model dependent
results for photonic parton distributions are considerably larger than those obtained perturbatively,
 even in the restricted kinematical region. In order to
illustrate this difference, in figure (2a) we show the ratio between the resolved 
and direct components of dijet
photoproduction imposing a threshold value of  $x_\gamma=1.0$ for a fixed value of $p_T$ ($p_T=4 \,GeV$), 
computed with different parametrizations.
The abbreviation `uem' stands for the perturbative parton distributions of 
reference \cite{uem}, `SaS 1D' and `SaS 2D' for those of \cite{ss}, `GRS'
for \cite{glu}, and `WHIT\#' for those of reference \cite{whit} with the implemented $P^2$ dependence
of reference \cite{DG2}.
Clearly, the perturbative distributions yield considerably smaller ratios, the main difference
 coming from the gluon content, as can be seen in figuere (2b)
where the perturbative result is compared with the quark component of the 
nonperturbative expectation for the ratio.\\

\noindent{\large \bf Experimental Data.}\\

In order to compare theoretical expectations with the available data on dijet photoproduction,
 it is necesary to find out the $x_{\gamma}$ threshold value
for the theoretical calculation that corresponds to the $x_{\gamma}^{obs}$ of
the experiment.
At $P^2=0.01\, GeV^2$ one would expect photons to behave almost as real photons
and the corresponding photonic parton distributions strongly constrained by the
real photon data. The model dependence implemented to take into account the 
virtuality, and the corresponding uncertainty, is then minimised so one can try
to find the value of $x_{\gamma}^{threshold}$ that provides the best accord 
between the data and the estimates.
In figure (3) we show the dijet ratios integrated in $p_T$ and at $P^2=0.01\, GeV^2$, as a function of 
$x_{\gamma}^{threshold}$  for different sets  against the experimental value
obtained with $x_{\gamma}^{obs}=0.75$ (the thick solid line). The conventions 
for the other lines are the same as in figures (2a) and (2b).
The figure shows that the best value for $x_{\gamma}^{threshold}$ lays between
$0.85$ and $0.95$. The residual uncertainty in the photonic parton distributions and the 
experimental errors prevents a more precise determination, however the values coming from most
 of the distributions are perfectly consistent with 
the definition of $x_{\gamma}^{obs}$, which implies $0.75\leq x_{\gamma}^{threshold} \leq 1.0$.
 The comparison also rules out some `extreme' 
distributions, such as the set WHIT4 \cite{whit}, which has an enormous gluon component and
would require $x_{\gamma}^{threshold}<<0.75$, or the perturbative set with 
almost null gluon content ($x_{\gamma}^{threshold}>>1$).

In figure (4) we compare the experimental data on dijet with the theoretical
expectations coming from different parametrizations as a function of $P^2$.
The lines correspond to the extreme estimates ($x_{\gamma}^{threshold}=1.0$ 
and $0.75$, as dashed lines) and to an intermediate value ($x_{\gamma}^{threshold}=0.85$, as a solid line).
The figures favor distributions with stronger dependence on the virtuality, 
such as SaS2D and GRS, both of which, incidentally, prefers
$x_{\gamma}^{threshold}\simeq 0.85$. No choice of 
$x_{\gamma}^{threshold}$ adjust the estimates coming from other paramentrizations 
to the data whith similar accuracy. Plausibly, future high statistics data on 
dijet photoproduction will allow a more stringent discrimination between parton
distributions.

In figure (5) we show estimates for single jet ratios. There, the low $x_{\gamma}$ contributions coming from the extra integration regions  increase 
the resolved component and thus the value for the ratio. However, the dependence on the virtuality scale and that on the $x_{\gamma}^{threshold}$,  seem to be
analogous to those of dijet ratios, implying a similar behaviour of each component inside and outside the restricted A'B'C' region. Apart form this, single jet data would not highlight individual characteristics of the sets, different to those shown by dijet data. However, it can help increasing the statistics of global fits. \\
\pagebreak

\noindent{\large \bf Conclusions:}\\

We have analysed different estimates for the ratio between direct and resolved 
contributions for single and dijet photoproduction. We have found a fairly good agreement between the estimates coming from most of the photonic parton distributions available and the data produced by ZEUS, and the impossibility to
conciliate these data with the estimates coming from  `extreme' parton
distributions, such as those of perturbative origin or with unlikely large gluon
content. The $P^2$-dependent data show greater agreement with estimates coming from parametrizations with stronger
$P^2$ dependence, a large gluon component, and a $x_{\gamma}^{threshold}$ value
of about $0.85$, corresponding to $x_{\gamma}^{obs.}=0.75$. Single jet ratios show slightly larger resolved components, but a similar $P^2$ dependence. Future
experiments on single jet and dijet photoproduction would be able to further constrain the different proposals for the parton structure in the virtual photon.    \\

\noindent{\large \bf Acknowledgements:}\\

We warmly acknowledge F. Barreiro for driving our attention to dijet photoproduction data and for kind comments and suggestions. 

\pagebreak

\pagebreak 

\noindent{\large \bf Figure Captions:}\\

\begin{itemize}
\item[{\bf Fig.1a}] Integration regions for the variables $\eta_{1}$ and $\eta_{2}$.
\item[{\bf Fig.1b}] Integration regions for the variables $\tilde{x}_{p}$ and 
$\tilde{x}_{\gamma}$.
\item[{\bf Fig.2a}] The ratio between the resolved and direct components of dijet photoproduction ($x_{\gamma}^{threshold}= 1.0$ and $p_T=4 \,GeV$), computed with different parametrizations.
\item[{\bf Fig.2b}]  The same as in Figure (2a) but only for the quark component
in the photon.
\item[{\bf Fig.3}] The dijet ratios integrated in $p_T$ and at $P^2=0.01\, GeV^2$ as a function of 
$x_{\gamma}^{threshold}$   and for different sets.  The experimental value
obtained with $x_{\gamma}^{obs}=0.75$ is also shown for comparison. 
\item[{\bf Fig.4}] The  data on dijet ratios against  the theoretical
expectations coming from different parametrizations as a function of $P^2$.
\item[{\bf Fig.5}] Estimates for single jet ratios as a function of $P^2$.
\end{itemize}      
 \vfill\eject
\begin{figure}[htb]
\begin{center}
\mbox{\kern-2cm
\epsfig{file=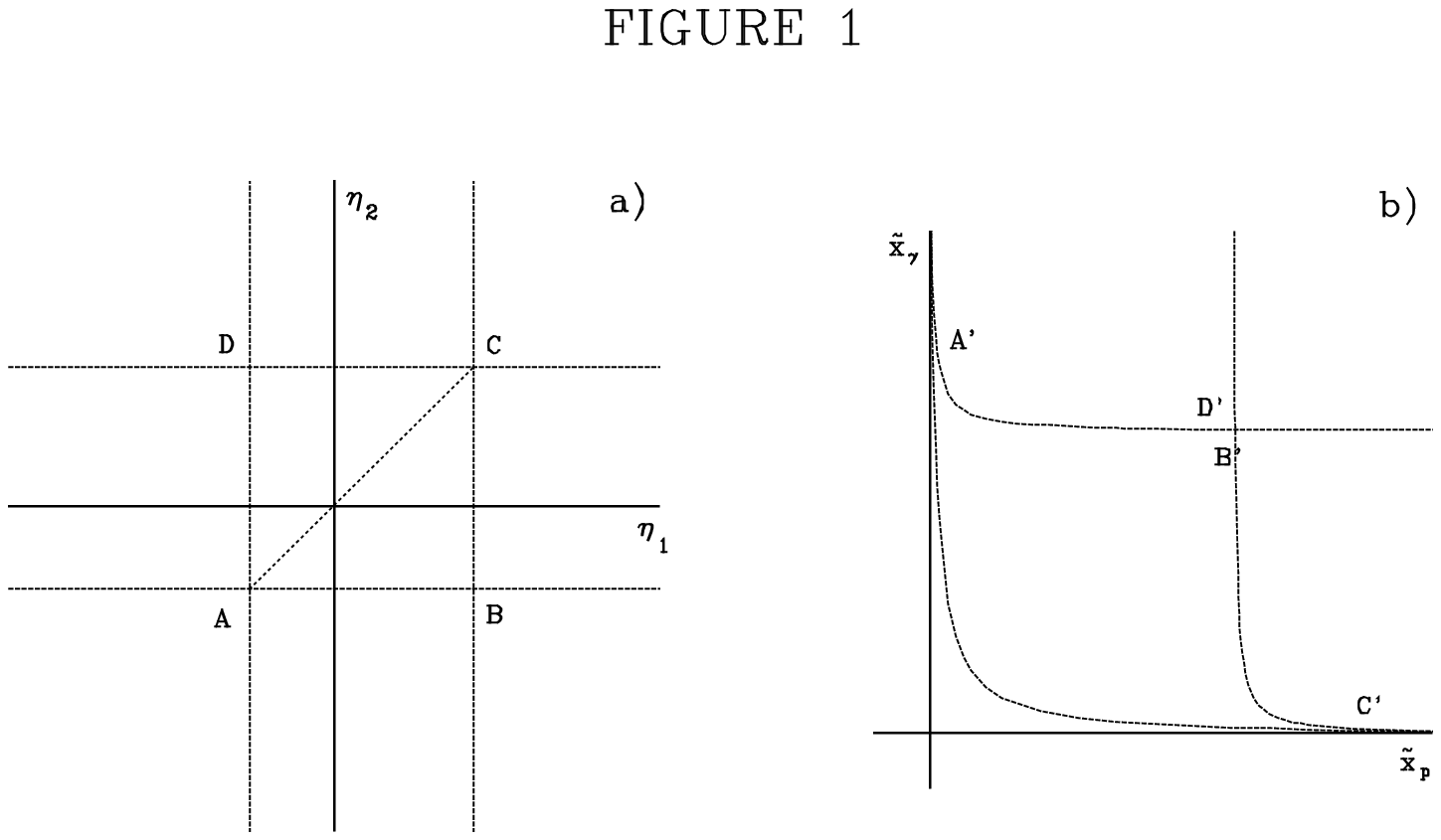,width=14.0truecm,angle=0}}
\caption{ a. Integration regions for the variables $\eta_{1}$ and $\eta_{2}$.
b. Integration regions for the variables $\tilde{x}_{p}$ and 
$\tilde{x}_{\gamma}$.}
\end{center}
\end{figure}
\vfill\eject

\begin{figure}[htb]
  \begin{center}
\mbox{\kern-2cm
\epsfig{file=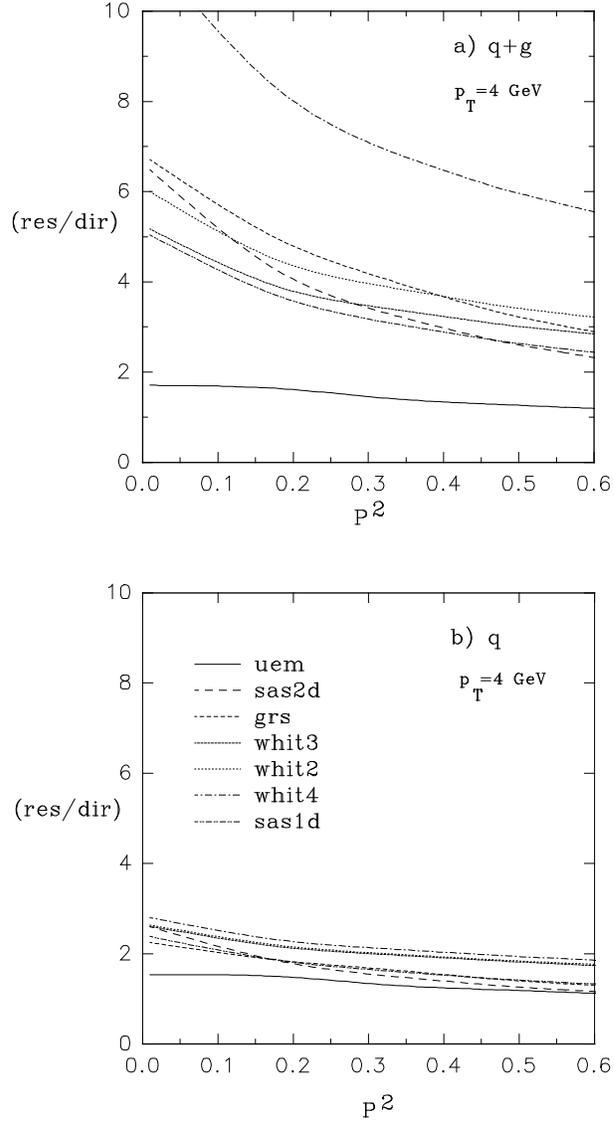,width=8.0truecm,angle=0}}
    \caption{a. The ratio between the resolved and direct components of dijet photoproduction ($x_{\gamma}^{threshold}= 1.0$ and $p_T=4 \,GeV$), computed with different parametrizations.
b. The same as in Figure (2a) but only for the quark component
in the photon.
}
  \end{center}
\end{figure}

\begin{figure}[htb]
  \begin{center}
\mbox{\kern-2cm
\epsfig{file=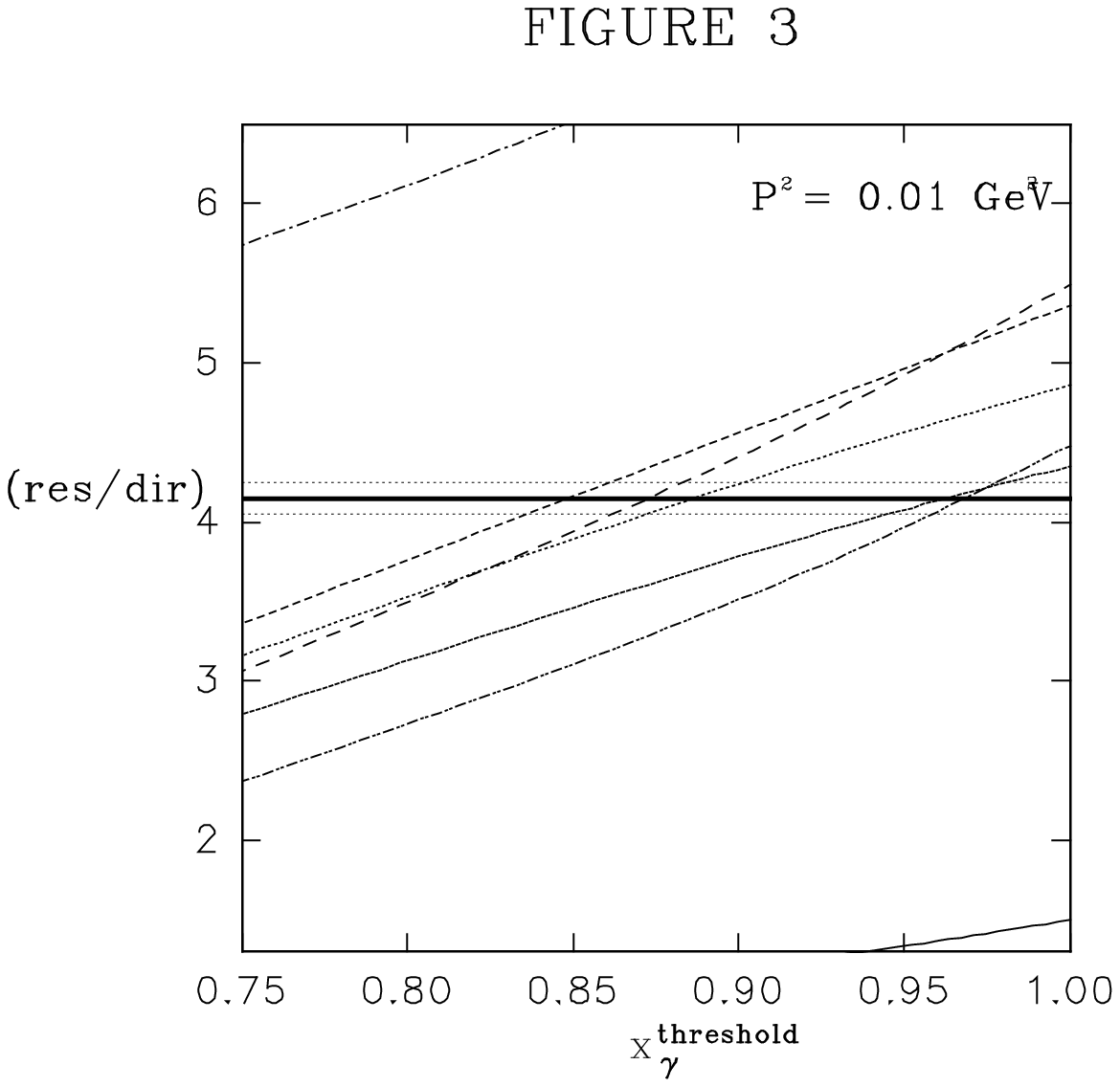,width=12.0truecm,angle=0}}
    \caption{ The dijet ratios integrated in $p_T$ and at $P^2=0.01\, GeV^2$ as a function of 
$x_{\gamma}^{threshold}$   and for different sets.  The experimental value
obtained with $x_{\gamma}^{obs}=0.75$ is also shown for comparison. 
}
  \end{center}
\end{figure}

\begin{figure}[htb]
  \begin{center}
\mbox{\kern-2cm
\epsfig{file=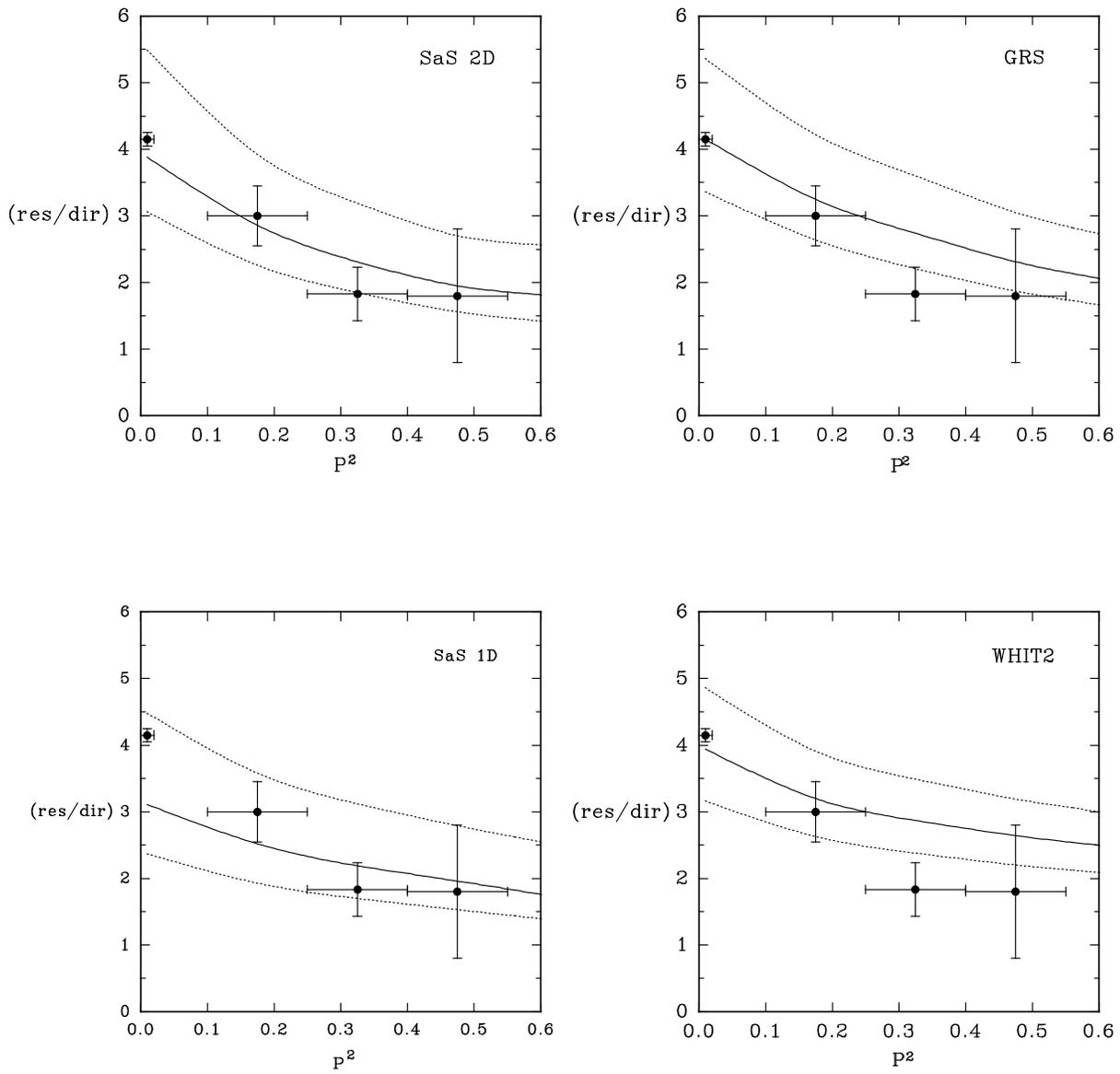,width=16.0truecm,angle=0}}
    \caption{The  data on dijet ratios against  the theoretical
expectations coming from different parametrizations as a function of $P^2$.
}
  \end{center}
\end{figure}
\vskip 14pt

\begin{figure}[htb]
  \begin{center}
\mbox{\kern-2cm
\epsfig{file=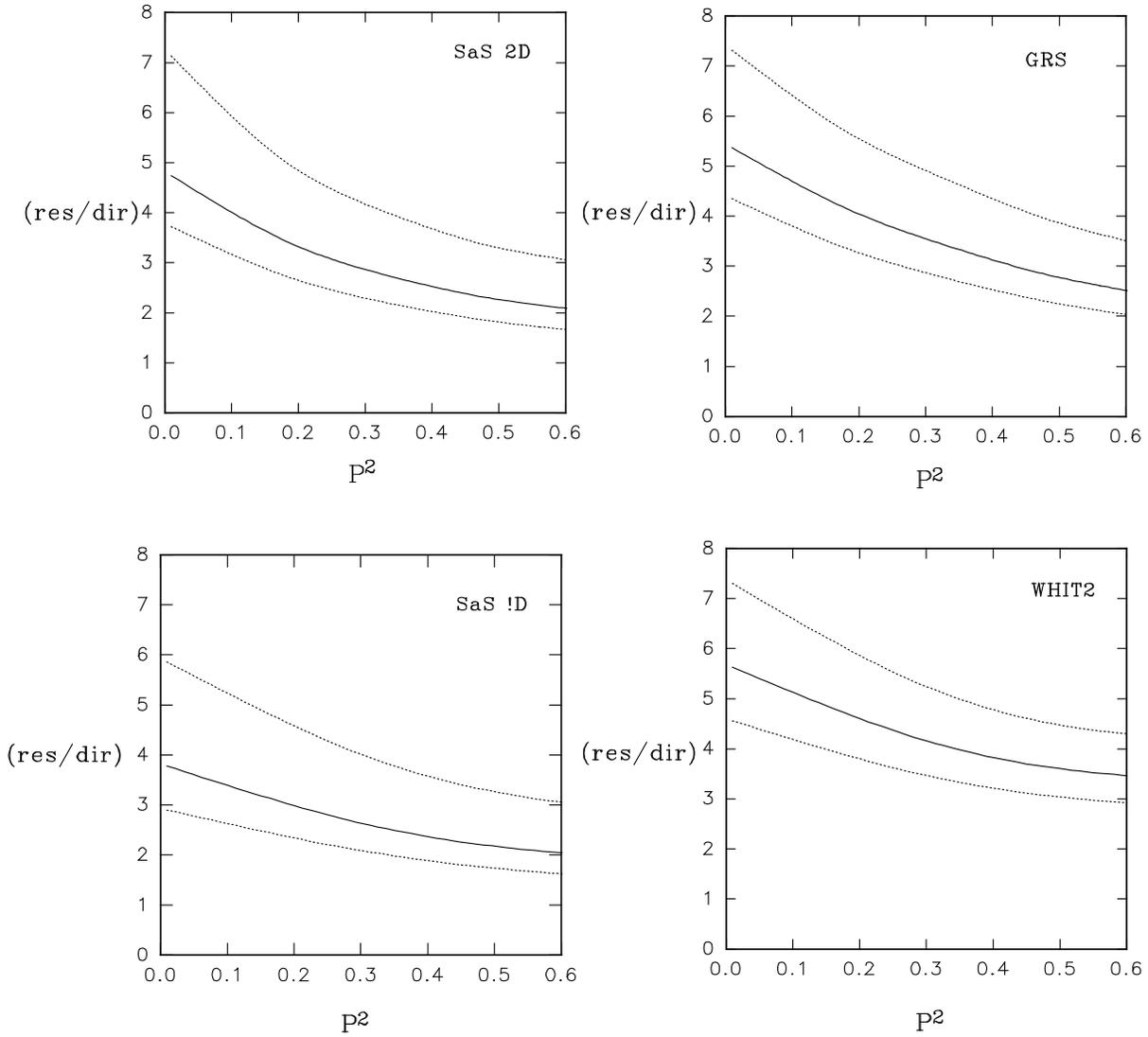,width=16.0truecm,angle=0}}
    \caption{ Estimates for single jet ratios as a function of $P^2$.
}
  \end{center}
\end{figure}

\vskip 14pt

\end{document}